\documentclass[twocolumn]{aa}
\usepackage{graphicx,amssymb}
\newcommand \msun {\mbox{$\mathcal{M}_{\odot}$}}
\newcommand \kms {\mbox{km~s$^{-1}$}}
\newcommand \degree {\mbox{$^\circ$}}

\begin{document}
   \title{Barred CO emission in HE~1029-1831\thanks{Based on
   observations carried out with the IRAM Plateau de Bure
   Interferometer and BIMA. IRAM is supported by INSU/CNRS (France),
   MPG (Germany) and IGN (Spain). The Berkeley-Illinois-Maryland
   Association (BIMA) observatory is supported by NSF grant AST
   99-81289.}}


   \author{M. Krips
          \inst{1,2}
          \and
          A. Eckart\inst{1}
	  \and
	  R. Neri\inst{3}
	  \and
	  T. Bertram\inst{1}
	  \and
	  C. Straubmeier\inst{1}
	  \and
	  S. Fischer\inst{1}
	  \and
	  J.G. Staguhn\inst{4}
 	  \and
          S.N. Vogel \inst{5}
         }

   \offprints{M. Krips\\ 
              e-mail: mkrips@cfa.harvard.edu}

   \institute{I. Physikalisches Institut, Universit\"at zu K\"oln,
              Z\"ulpicher Str. 77, 50937 K\"oln
              \and
	      Harvard-Smithsonian Center for Astrophysics, SMA
              project, 645 North A`ohoku Place, Hilo, HI 96720;
              \email{mkrips@cfa.harvard.edu}    
              \and
              Institut de Radio Astronomie Millimetrique (IRAM), 300 rue de 
              la Piscine, 38406 Saint Martin d'H\`eres, France
              \and
              NASA/Goddard Space Flight Center, Code 665, Building 21, 
	      Greenbelt, MD 20771, USA
              \and      
	      University of Maryland, College Park, MD 20742, USA             }
   \date{Received ; accepted }

   \abstract{We present CO(1--0) and CO(2--1) line emission maps of
     the barred spiral active galaxy HE~1029-1831 (z=0.0403)
     obtained with the IRAM Plateau de Bure Interferometer (PdBI) and
     in part by the Berkeley-Illinois-Maryland Association (BIMA)
     observatory. The CO emission is well associated with the optical
     bar and extended along it. The FWHM of the CO emission is
     estimated to be $\sim$(6$\pm$2)~kpc. The CO emission shows a
     strong velocity gradient along the minor axis of the bar
     (PA=90\degree). The molecular gas mass is estimated to be
     $\sim$1.2$\times10^{10}$\msun\, which indicates a very gas rich
     host galaxy. Most of the molecular gas appears to be subthermally
     excited and cold but we also find weak evidence for a warmer
     and/or denser gas component at the southern part of the bar
     emission, about $\sim$4~kpc from the galactic
     nucleus.  }

   \maketitle
%

\section{Introduction}
HE~1029-1831 is part of a nearby QSO sample (Bertram et al., in prep.)
consisting of $\sim$100 sources from the Hamburg/ESO survey for bright
QSOs. The sole selection criterion of these sources is their low
redshift of $z<0.06$ providing the capability of observations on the
smallest observationally available angular scales and the
accessibility of several important diagnostic lines in the NIR (e.g.,
the CO(2--0) rotation vibrational band head absorption line). We aim at
determining the distribution and dynamics of molecular gas in the
inner 1~kpc of these objects. This will add to a systematic study of
the different mechanisms of gas fueling into active galactic nuclei
(AGN). Together with the NUGA PdBI survey containing Seyfert and LINER
galaxies with z$\leq$0.013 (Garc\'{\i}a-Burillo et al.\ 2003a,b,2005;
Combes et al.\ 2004; Krips et al., 2005) this study will cover a large
part of the activity sequence of AGN. Especially, the role of nuclear
embedded bars, rings, spirals as well as mini-spirals, warps and
micro-warps in the fueling process is an important issue to
investigate possible links between different nuclear disk morphologies
and activity types.

Thirty objects of this sample were observed in CO with the SEST
telescope and ten with the BIMA array (Bertram et al., in prep.)
resulting in a low detection rate of\, $\sim$15\% which might be a
result of a lack of instrumental sensitivity and uncertainties in the
given redshifts; these two observing campaigns were initially
conducted to identify the objects with the strongest CO emission. As a
consequence of the low detection rate, we have carried out deeper
integration of an additional 26 sources with the IRAM 30m telescope
leading to 19 further detections (Betram et al.\ 2006, Bertram et
al.\, in prep.). The BIMA observations resulted in a clear detection
of the source HE~1029-1831 and in a tentative one of
HE~1136-2304. Both have thus been observed as a follow-up with the
IRAM PdBI array but only the first one could be detected and mapped
while the tentative BIMA detection of HE~1136-2304 could not be
confirmed. HE~1136-2304 most likely lacks an exact redshift, too.

\begin{figure}[!t]
\centering
\resizebox{\hsize}{!}{\rotatebox{-90}{\includegraphics{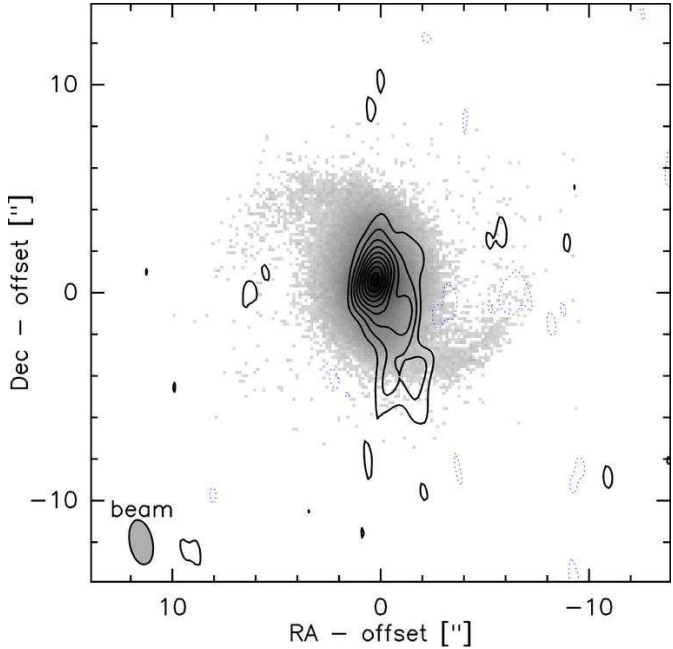}}}
\caption{H-band image of HE1029-1831 (from Fischer et al.\ 2006)
superimposed with contours of the integrated CO(2--1) line
emission. Contours are from 3$\sigma=$1.4\ Jy/beam\ \kms to 13\
Jy/beam\ \kms in steps of $3\sigma$.}
\label{he1029-opt-co}
\end{figure}

HE~1029-1831 contains (m=2) spiral arms and a prominent bar
(PA$\simeq$30\degree) in the optical (H-band; Fischer et al.\ 2006;
see Fig.~\ref{he1029-opt-co}). Its redshift was recently determined to
be $z=0.0403\pm0.0001$ (Kaldare et al.\ 2003).  The classifications of
its active nucleus range from an HII region and extreme starburst to
an AGN (Kewley et al.\ 2001). However, recent NIR observations have
given evidence that HE~1029-1831 may be classified as a luminous
infrared galaxy with a weak narrow line Seyfert nucleus of type 1
(Fischer et al., 2006).

\begin{figure}[!t]
\centering
\resizebox{\hsize}{!}{\rotatebox{-90}{\includegraphics{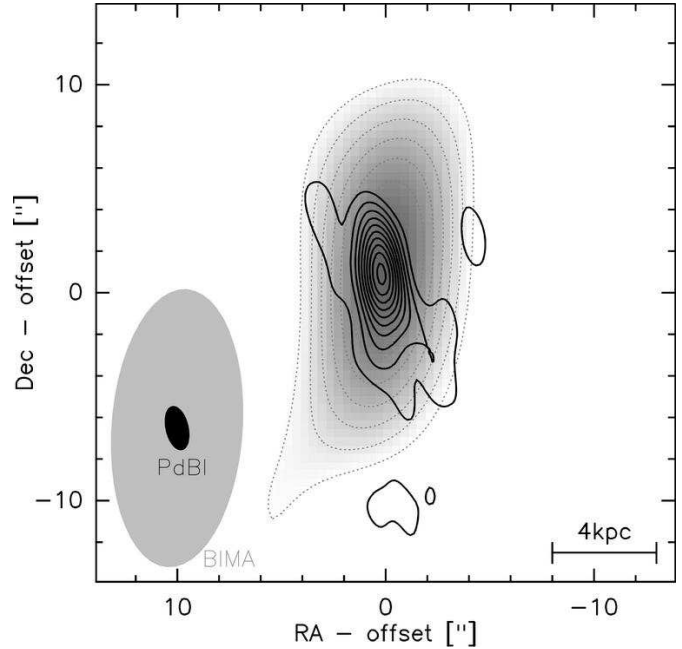}}}
\caption{Integrated CO(1--0) emission as observed with BIMA ({\it grey
scale + dotted grey contours}). The PdBI CO(1--0) ({\it solid black
contours}) are superimposed on the image. The contour lines for the
BIMA data represent multiples of 1$\sigma$ levels, starting at
3$\sigma$, where $1\sigma$=3.0 Jy/beam\ \kms. The contours for the
PdBI CO(1--0) emission run from 3$\sigma=$0.8\ Jy/beam\ \kms\ to 7.3\
Jy/beam\ \kms\ in steps of $3\sigma$.}
\label{he1029-bima}
\end{figure}

\begin{table*}[!t]
\centering
\begin{tabular}{ccccccc}
\hline
\hline
Line &  line & line & integrated &  integrated intensity &  z & 
M(H$_2$+He)$^b$\\
& peak flux$^f$ & width &  peak intensity & in central $\sim$20$''$ 
&   & \\
 & [mJy/beam] & [\kms] & [Jy/beam\,\kms] & [Jy\ \kms] &
& [10$^9$\msun]\\
\hline
\multicolumn{7}{c}{PdBI}\\
\hline
CO(1--0) & 65$\pm$7 & 121$\pm$3 & 8.4$\pm$0.2 & 21$\pm$1  
& 0.0402$\pm$0.0001 & 1-8 \\
CO(2--1) & 90$\pm$20 & 131$\pm$7 & 12$\pm$0.5 & 25$\pm$1
& 0.0402$\pm$0.0001 & - \\
\hline
\multicolumn{7}{c}{BIMA}\\
\hline
CO(1--0) & 190$\pm$30 & 150$\pm$20 & 28$\pm$3 & 30$\pm$3  
& 0.0402$\pm$0.0002 & 2-12 \\
\hline
\hline
 & \multicolumn{2}{c}{Centre of CO emission} & 
\multicolumn{2}{c}{Dynamical Centre of CO emission} &  PA$^c$ & i$^d$  \\
& RA$^e$ & Dec$^e$ &  RA$^e$ & Dec$^e$ & [\degree] & [\degree] \\
\hline
\multicolumn{7}{c}{PdBI}\\
\hline
CO(1--0) & 10$^{\rm h}$31$^{\rm m}$57.3$^{\rm s}$ & $-$18\degree46$'$33.2$''$  &
10$^{\rm h}$31$^{\rm m}$57.3$^{\rm s}$ & $-$18\degree46$'$33.4$''$ & $\sim$20
& $\sim$10-30 \\
CO(2--1) & 10$^{\rm h}$31$^{\rm m}$57.3$^{\rm s}$ & $-$18\degree46$'$33.2$''$  &
 10$^{\rm h}$31$^{\rm m}$57.3$^{\rm s}$ & $-$18\degree46$'$33.4$''$ & $\sim$3
& - \\
\hline
\multicolumn{7}{c}{BIMA}\\
\hline
CO(1--0) & 10$^{\rm h}$31$^{\rm m}$57.3$^{\rm s}$ & $-$18\degree46$'$33.8$''$  &
10$^{\rm h}$31$^{\rm m}$57.3$^{\rm s}$ & $-$18\degree46$'$33.8$''$ & $\sim$0 
& $\sim$10-30 \\
\hline
\end{tabular}
\caption{Line parameters for CO(1--0) and CO(2--1) at the peak and
integrated over the central 10-20$''$. $^a$ taking a luminosity
distance of 180~Mpc (based on the following cosmology: $H_0=70$\kms,
$\Omega_m$=0.3 and $\Omega_\lambda$=0.7; 1$''$ corresponds then to
$\sim 0.8$kpc). $^b$ assuming a M(H$_2$) to $L'_{\rm CO}$ conversion
factor of 0.8-4.8~\msun\ (K\kms\ pc$^2$)$^{-1}$ (Solomon \& Barrett
1991; Downes \& Solomon 1998) and
M(H$_2$+He)=1.36$\times$M(H$_2$). $^c$ Estimated position angle of the
CO emission (from North to East). $^d$ Estimated inclination angle of
the CO disk. $^e$ The position uncertainties are of the order of
$\sim$0.2$''$ for the PdBI data and of $\sim$0.4$''$ for the BIMA
data. $^f$ flux errors include $\sim$10\% ($\sim$20\%) uncertainties
from the flux calibration at 3~mm (1~mm).}
\label{he1029-linepara}
\end{table*}

\begin{figure}[!]
\centering
\resizebox{\hsize}{!}{\rotatebox{-90}{\includegraphics{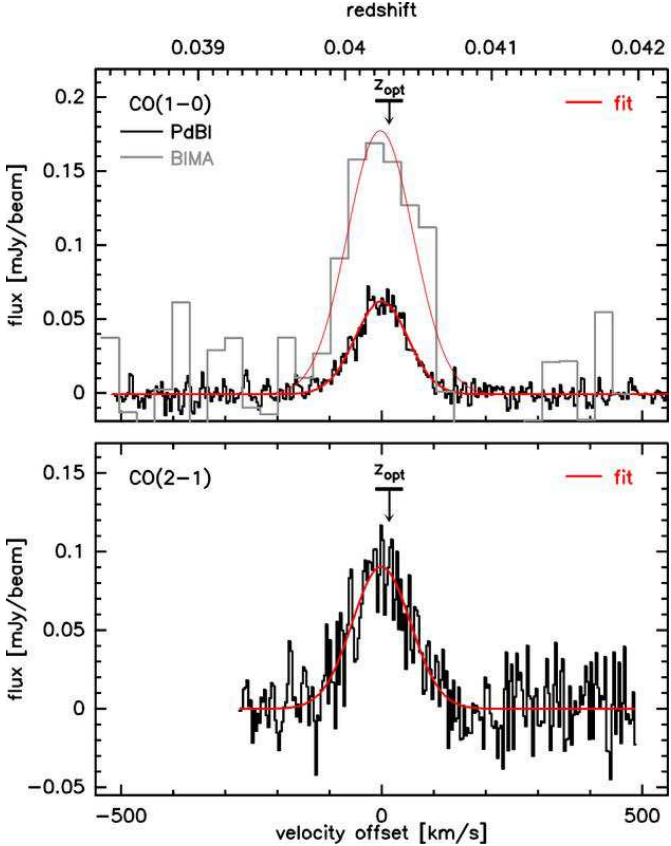}}}
\caption{CO(1--0) ({\it upper panel}) and CO(2--1) ({\it lower panel})
line spectrum taken at the peak emission. The (optical) redshift of
z=0.0403$\pm$0.0001 is indicated. The velocity resolution is 3.4\kms\
(34~\kms) at both lines for the PdBI (BIMA).}
\label{he1029-spe-co}
\end{figure}

\section{Observations}

\subsection{BIMA}
HE~1029-1831 is one of 10 members of the sample that were scanned for
CO(1$-$0) emission with BIMA at Hat Creek, CA, USA between October
2000 and April 2001.  The interferometer, consisting of 10 antennae
with a diameter of 6.1m each, was used in C configuration.  The
correlator was configured to provide 8 spectral windows in the upper
sideband with 32 channels each and covered an overall bandwidth of
800~MHz.  The observations were scheduled over a range of several
months with varying weather conditions, total integration time and
uv-coverage for the different sources, hence resulting in different
spatial resolutions and detection limits. In the case of HE~1029-1831,
the BIMA survey led to a detection of bright CO line emission
(Fig.\ref{he1029-bima}).  These data were reduced using 3C273 as
bandpass and 1048-191 as gain calibrator. A medium spatial resolution
of 13.6\arcsec$\times$5.8\arcsec\, was achieved in this antenna array
configuration.

\subsection{PdBI}
As BIMA follow-up, observations of the CO(1--0) and CO(2--1) line
emission in HE~1029-1831 were conducted in February 2002 with 6
antennae in B and C configurations of the IRAM PdBI and again in March
2003 with 6 antennae in A configuration.  The bandpass calibrator was
set to 3C273 and amplitude and phase calibrators were 1048-191 and
1055+018. The latter were observed every twenty minutes. The 3.5~mm
receiver was tuned to the redshifted $^{12}$CO(1--0) line and the
1.2~mm receiver to the redshifted $^{12}$CO(2--1) line.  The amount of
water vapor is calculated to be $\leq$6~mm for both sets of
observations thus providing reasonable weather conditions and system
temperatures for a low declination target. At both frequencies we used
a total bandwidth of 580~MHz with a frequency resolution of
1.25~MHz. The phase tracking centre was set to
$\alpha_{2000}=$10$^{\rm h}$31$^{\rm m}$57.3$^{\rm s}$ and
$\delta_{2000}$=-18\degree46$'$34.0$''$ taken from the Nasa/IPAC
Extragalactic Database (NED). The total integration time on source (6
antennae) amounted to $\sim$5~hours for the 2002 data and to
$\sim$3~hours for the 2003 data.

\section{The data - results}
Both, the CO(1--0) and CO(2--1) lines are clearly detected
(Fig.~\ref{he1029-bima}, \ref{he1029-spe-co} and \ref{he1029-int-co})
while only upper limits can be determined for the continuum emission
at 3~mm and 1~mm.

\subsection{CO emission}
\begin{figure}[!t]
\centering
\resizebox{\hsize}{!}{\rotatebox{-90}{\includegraphics{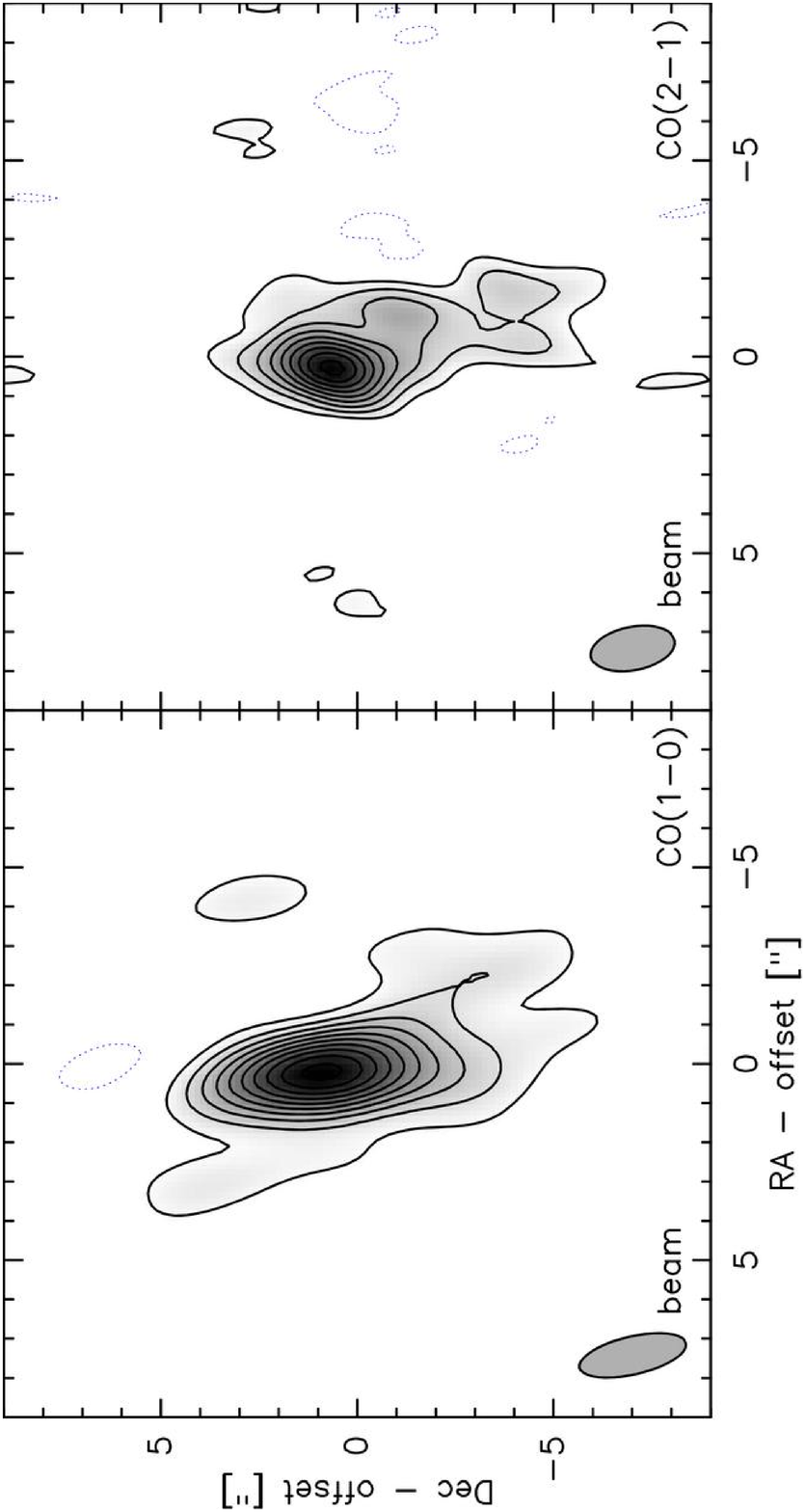}}}
\resizebox{\hsize}{!}{\rotatebox{-90}{\includegraphics{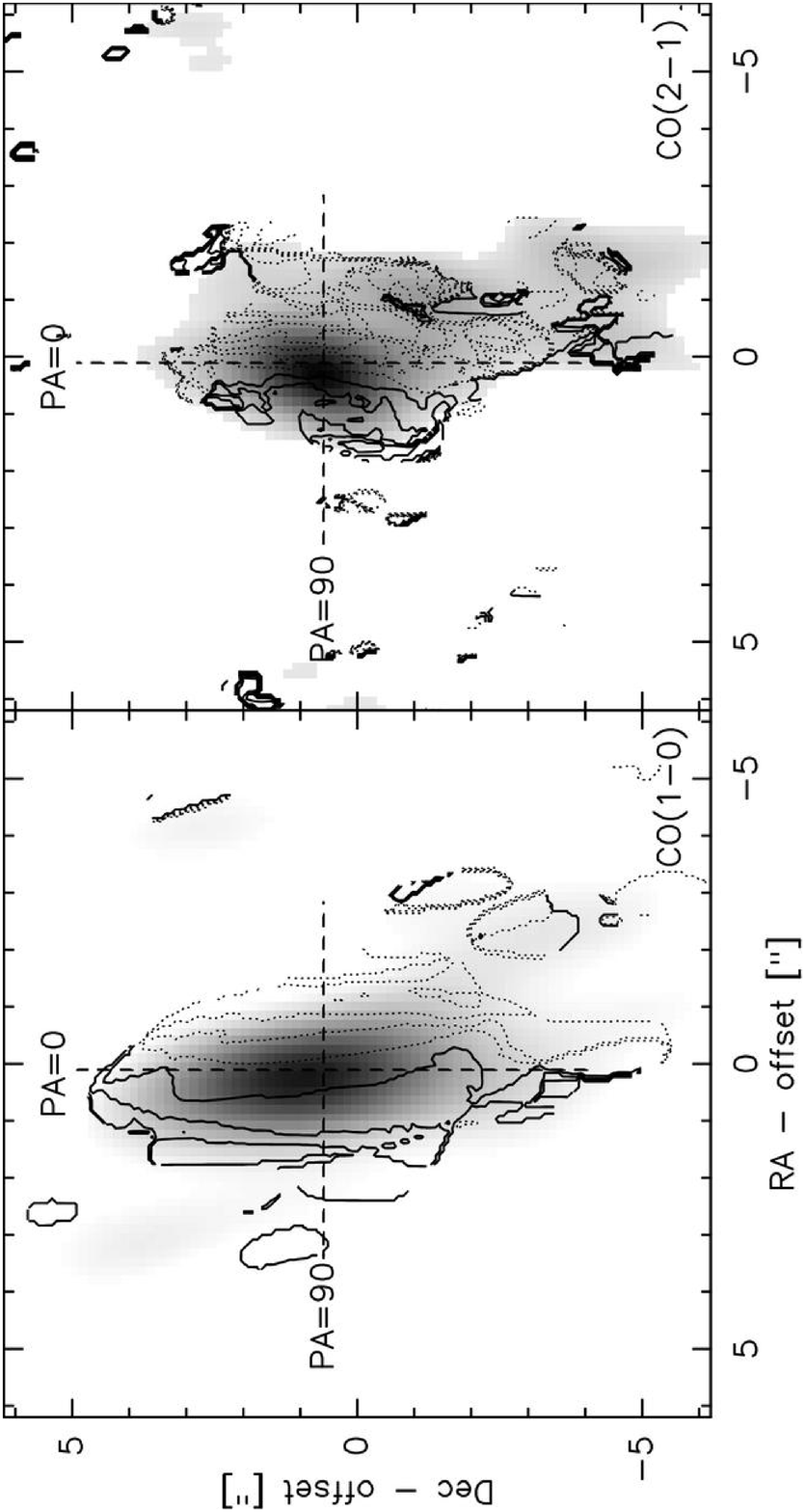}}}
\caption{{\it Upper panel:} Velocity integrated CO(1--0) ({\it left})
and CO(2--1) ({\it right}) emission in HE~1029-1831 (between
$\pm$200\kms). Contours are from 3$\sigma=$0.8\ Jy/beam\ \kms\ (1.4\
Jy/beam\ \kms) to 7.3\ Jy/beam\ \kms\ (13\ Jy/beam\ \kms) in steps of
$3\sigma$ for CO(1--0) (CO(2--1)). {\it Lower panel:} Iso-velocity
maps of the CO(1--0) and CO(2--1) emission CO(1--0) ({\it left}) and
CO(2--1) ({\it right}) emission in HE~1029-1831. Contours are around
the dynamic centre in steps of 10~\kms. The dashed lines indicate the
cuts along which the position-velocity diagrams in
Fig.~\ref{he1029-sli-co} were taken. }
\label{he1029-int-co}
\end{figure}

CO(1--0) emission was first detected with the BIMA array
(Fig.~\ref{he1029-bima}) and confirmed in subsequent observations of
CO(1--0) and CO(2--1) with the IRAM PdBI. Fig.\ref{he1029-spe-co}
shows the spectrum of the CO(1--0) and CO(2--1) line taken at the
centroid of the CO emission for the IRAM PdBI and BIMA data. The
redshift of the CO emission is determined to be z=0.0402$\pm$0.0001
for both transitions and both arrays which is consistent with the
optical redshift of z=0.0403$\pm$0.0001 (Fig.~\ref{he1029-spe-co};
Kaldare et al.\ 2003). A gaussian profile has been fitted to both
lines. The calculated line parameters are given in
Table~\ref{he1029-linepara}. The integrated maps are plotted in
Fig.~\ref{he1029-bima} \& \ref{he1029-int-co}. Both CO lines are
centered at the same position of $\alpha_{\rm J2000}=$10:31:57.3 and
$\delta_{\rm J2000}=-$18:46:33.2 (Table~\ref{he1029-linepara}). The
CO(1--0) emission is not resolved by the modest angular resolution
($\leq$14$''$) of the BIMA observations; the velocity integrated flux
at peak and in the inner $\sim$20$''$ are almost identical (see
Table~\ref{he1029-linepara}). This excludes the existence of
significant, very large scale ($>$10kpc) CO emission. The CO emission
is, however, clearly extended in both transitions at the higher
angular resolution (1-3$''$) of the PdBI and reveals elongations in
north to south direction with a position angle (PA) of
$\sim$0-20\degree (see Table~\ref{he1029-linepara}) which is slightly
smaller than the one derived for the optical bar (PA$\simeq$30\degree;
see Fischer et al.\ 2006). A combination of the BIMA and IRAM PdBI
CO(1--0) maps enables an estimate of the CO size ($\equiv$FWHM) by
fitting an elliptical component to the uv data. We find a FWHM of
$\sim$(7$\pm$2)$''$ which corresponds to $\sim$(6$\pm$2)~kpc. The
integrated CO(1--0) line intensity results in a total gas mass of
M(H$_2$+He)$\approx$8$\times10^9$\msun\, for the PdBI map and of
$\sim$1.2$\times10^{10}$\msun\, for the lower angular resolution BIMA
map assuming a M(H$_2$) to $L'_{\rm CO}$ conversion factor of
4.8~\msun\ (K\kms\ pc$^2$)$^{-1}$ (Solomon \& Barrett 1991) and
M(H$_2$+He)=1.36$\times$M(H$_2$). The difference in integrated
intensity between the PdBI and BIMA maps indicates that the PdBI has
filtered out approximately 30\% of the total CO emission due to the
lack of short-spacings. Taking the IRAS flux of 3.7~Jy at 3~THz (taken
from NED) and assuming optically thin dust emission with a
(rest-frame) dust temperature\footnote{This value was estimated by
fitting a grey body spectrum to the IRAS fluxes published by the
NED. } of $\sim$40~K, we obtain a dust mass of
$\sim$1$\times$10$^7$\msun. For a standard galactic gas-to-dust mass
ratio of 150, the expected gas mass would be
$\sim$2$\times$10$^9$\msun, i.e., a factor of $\sim$4-6 lower than
using the CO luminosity. However, the assumed gas-to-dust mass ratio
contains a large uncertainty and can increase in the cases of a few
active Ultra-Luminous InfraRed Galaxies (ULIRGs) even up to 1000 or
more (e.g., Contini \& Contini 2003) so that the galactic value might
not be applicable for HE~1029-1831. Alternatively, Downes \& Solomon
(1998) pointed out that the M(H$_2$) to $L'_{\rm CO}$ conversion
factor might be lower by a factor of 6 in the circumnuclear regions of
ULIRGs.

\begin{figure}[!t]
\centering
\resizebox{\hsize}{!}{\rotatebox{-90}{\includegraphics{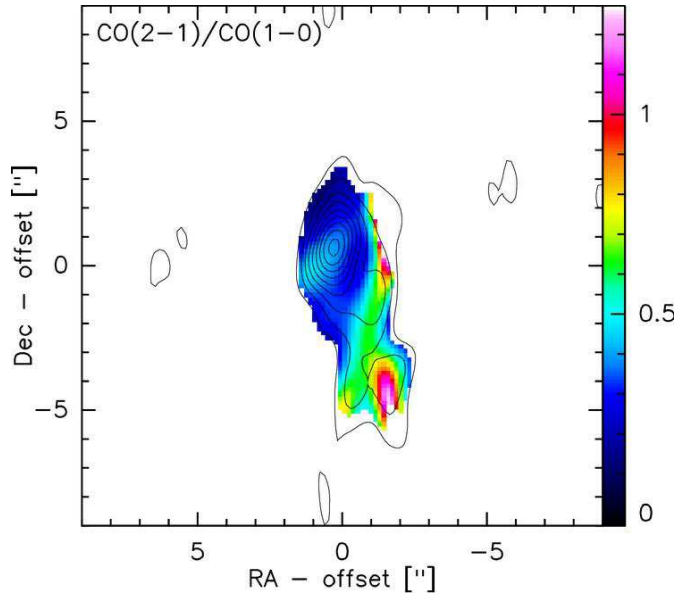}}}
\caption{The CO(2--1)/CO(1--0) line ratio in temperature scale ({\it
color scale}) overlaid with the velocity integrated CO(2--1) emission
({\it contours}). Increments for the velocity integrated CO(2--1)
emission are the same as in Fig.~\ref{he1029-int-co}.}
\label{he1029-co-line}
\end{figure}

As the angular resolutions of the PdBI for both transitions are
comparable when using uniform weighting for CO(1--0) and natural
weighting for CO(2--1), we can estimate the central ($\leq$8$''$) line
ratio $R_{21}$=CO(2--1)/CO(1--0) (in temperature scale; i.e.,
$R_{21}=$(S$_{\rm CO21}$(Jy)$\cdot\nu_{\rm obs,CO21}^{-2}$)/(S$_{\rm
CO10)}$(Jy)$\cdot\nu_{\rm obs,CO10}^{-2}$) for equal beamsizes)
between both lines to be $\simeq$0.5, indicating subthermal excitation
conditions and cold gas. However, one has to keep in mind that this
estimate might still be biased by artifacts due to different
$uv$-coverages for both lines in terms of spatial frequencies. One can
reduce such a bias by apodising and truncating the respective
$uv$-coverages to the overlapping region.  This results in the map of
line temperature ratios shown in Fig 5.  The ratio map in this figure
is not significantly different from one obtained simply from the
unchanged uniformly weighted CO (1-0) and naturally weighted CO (2-1)
maps. However, the resulting line ratio map still has to be
interpreted with caution as it heavily depends on the adopted
weighting factors. Thus, the integrated line ratio must be regarded as
a best estimate. Although most of the CO appears to be subthermally
excited and cold, the CO(2--1) line also indicates asymmetric emission
at the south which has only a very weak counterpart in the CO(1--0)
line. Such a difference cannot be explained by resolution effects but
could still be accounted for by a residual calibration error in the
visibilities and a poor coverage of the uv-plane although no obvious
inconsistencies were found. The high line ratio at the southern part
of $\sim$1 might thus indicate different excitation conditions of the
gas. The location of the southern CO component is close to the
southern spiral arm seen in the optical (see
Fig.~\ref{he1029-opt-co}). This might suggest that the CO at this
position traces the contact point of the bar and the southern spiral
arm. The high line ratio could be then indicative of gas compression
caused by a crossing of the bar- with the spiral arm orbits outside
the bar corotation, consistent with standard theory. However, the
southern CO emission has to be verified by further observations using
higher transitions of the CO line or denser gas tracers such as
$^{13}$CO or HCN.

\begin{figure}[!t]
\centering
\resizebox{\hsize}{!}{\rotatebox{-90}{\includegraphics{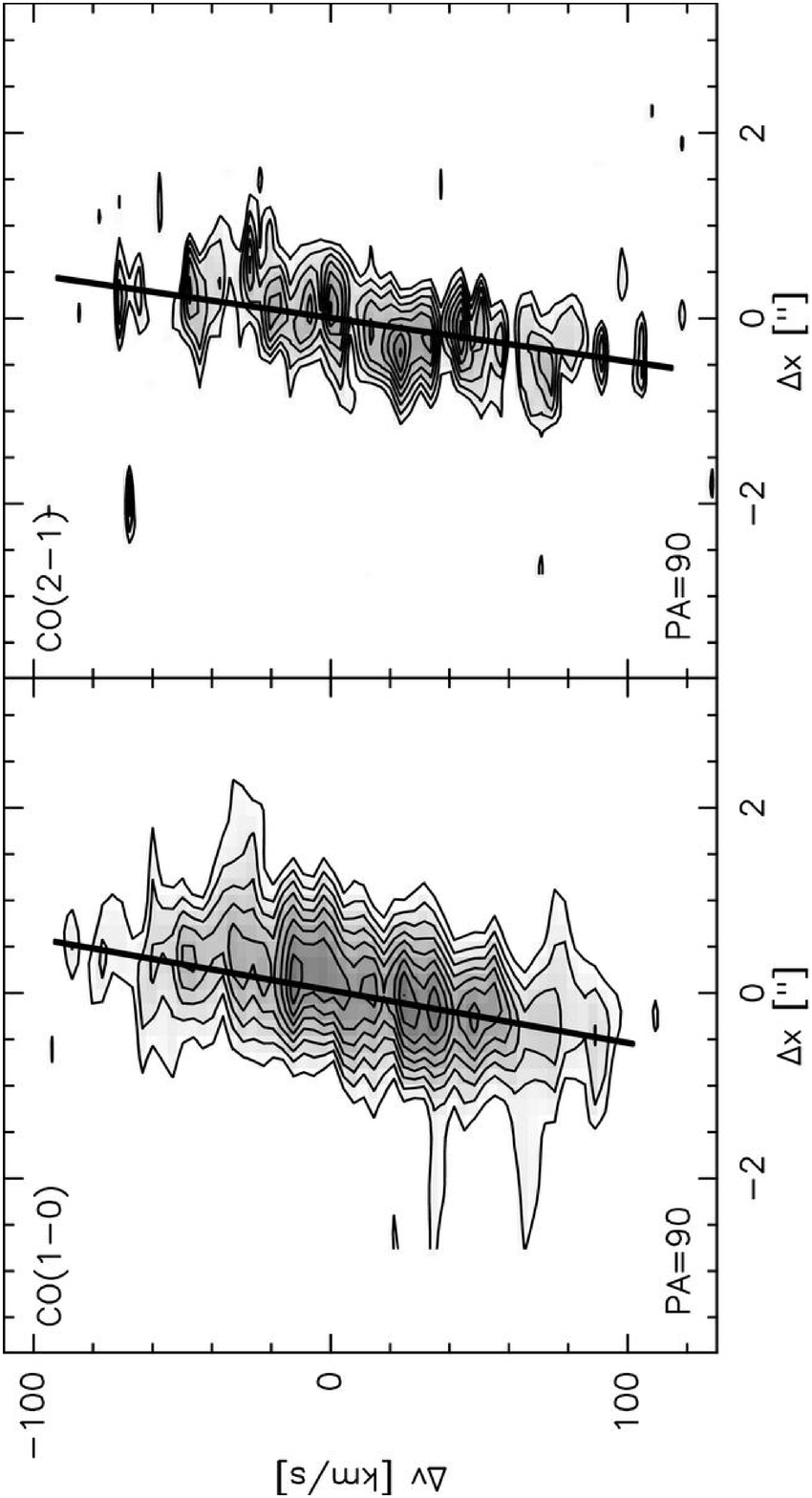}}}
\resizebox{\hsize}{!}{\rotatebox{-90}{\includegraphics{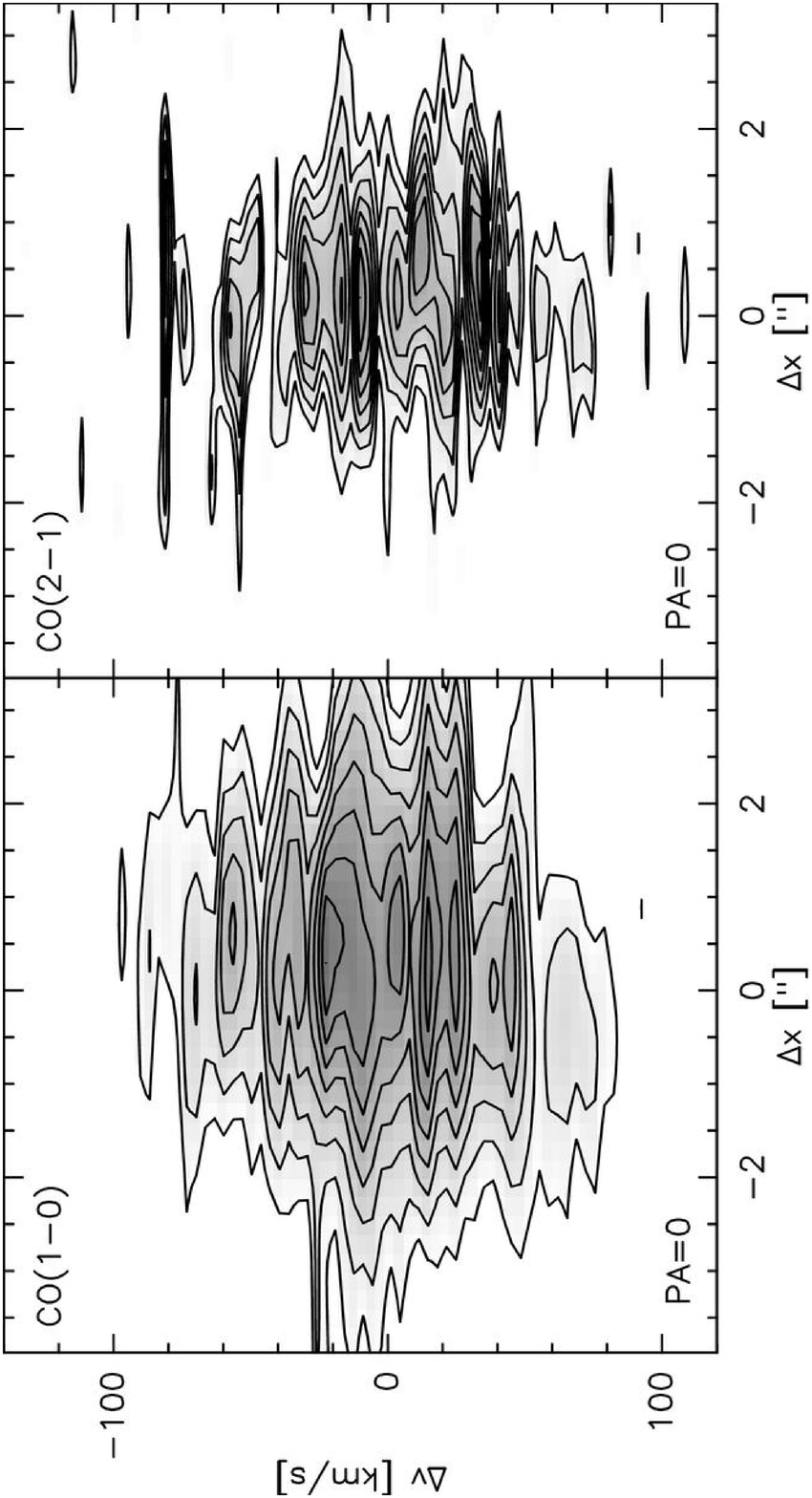}}}
\caption{The upper panels show the position velocity diagrams taken
  across the bar (PA=90\degree), and the lower panel those of a cut
  perpendicular to it, i.e., along the bar (PA=0\degree; see
  Fig.~\ref{he1029-int-co}). Increments are in steps of 10\% from 20\%
  and 30\% of the maximum for the CO(1--0) and the CO(2--1) line
  respectively. The thick lines indicate the steepness of the gradient
  at PA=90\degree\, and are supposed to guide the reader's eyes.}
\label{he1029-sli-co}
\end{figure}

We have taken position velocity diagrams at all position angles with a
separation of 10\degree\, showing that the cuts in the direction of
the CO extension (position angle: PA=0\degree) and perpendicular to it
(PA=90\degree) represent the extremes in velocity structure: the cuts
at PA=90\degree\, in both transitions reveal a clear and strong
velocity gradient (Fig.~\ref{he1029-sli-co}, upper panels) which
extends over a positional range of $\pm$0.5$''$; by comparison,
velocity gradients along the CO elongation, i.e., at PA=0\degree, are
weaker and much less apparent (Fig.~\ref{he1029-sli-co}, lower
panels). Assuming a radius of $\sim1''$ equivalent to 0.8~kpc and a
velocity of $\sim80$\kms, the total dynamical mass within the central
0.8~kpc - not corrected for inclination effects - can be calculated to
be $\geq2\times10^9$\msun. Probably, the gas disk is seen closer to
face-on than to edge-on (compare also Fischer et al.\ 2006) so that
this value must be regarded as lower limit. Since the dynamical mass
includes the mass of the gas {\it and} of the stars, we can estimate
the inclination of the galaxy by assuming that the gas mass of
$\sim0.2-1.2\times10^{10}$\msun\, contributes 10\% to the total mass
of the galaxy. This results in an inclination of $\sim$5-10\degree. By
equating directly the estimated gas mass with the dynamical mass, the
inclination would increase to $<$15-30\degree\, giving a more
conservative upper limit.

\subsection{Continuum emission}
The continuum emission at 3~mm and 1~mm was not detected. The upper
limits were determined in the line free channels (i.e., at velocities
lower than $-250$\kms\ and larger than $+250$\kms\ at 3~mm and at
velocities lower than $-200$\kms\ and larger than $-200$\kms\ at 1~mm)
to $3\sigma$=3~mJy at 3~mm and to $3\sigma=8$~mJy at 1~mm. The two
upper limits are compatible with the mm-values predicted by the grey
body spectrum of the (infrared) dust emission (compare previous
Section) which are of the order of $\sim$1-5~mJy.

\begin{figure}[!]
\centering
\resizebox{\hsize}{!}{\rotatebox{-90}{\includegraphics{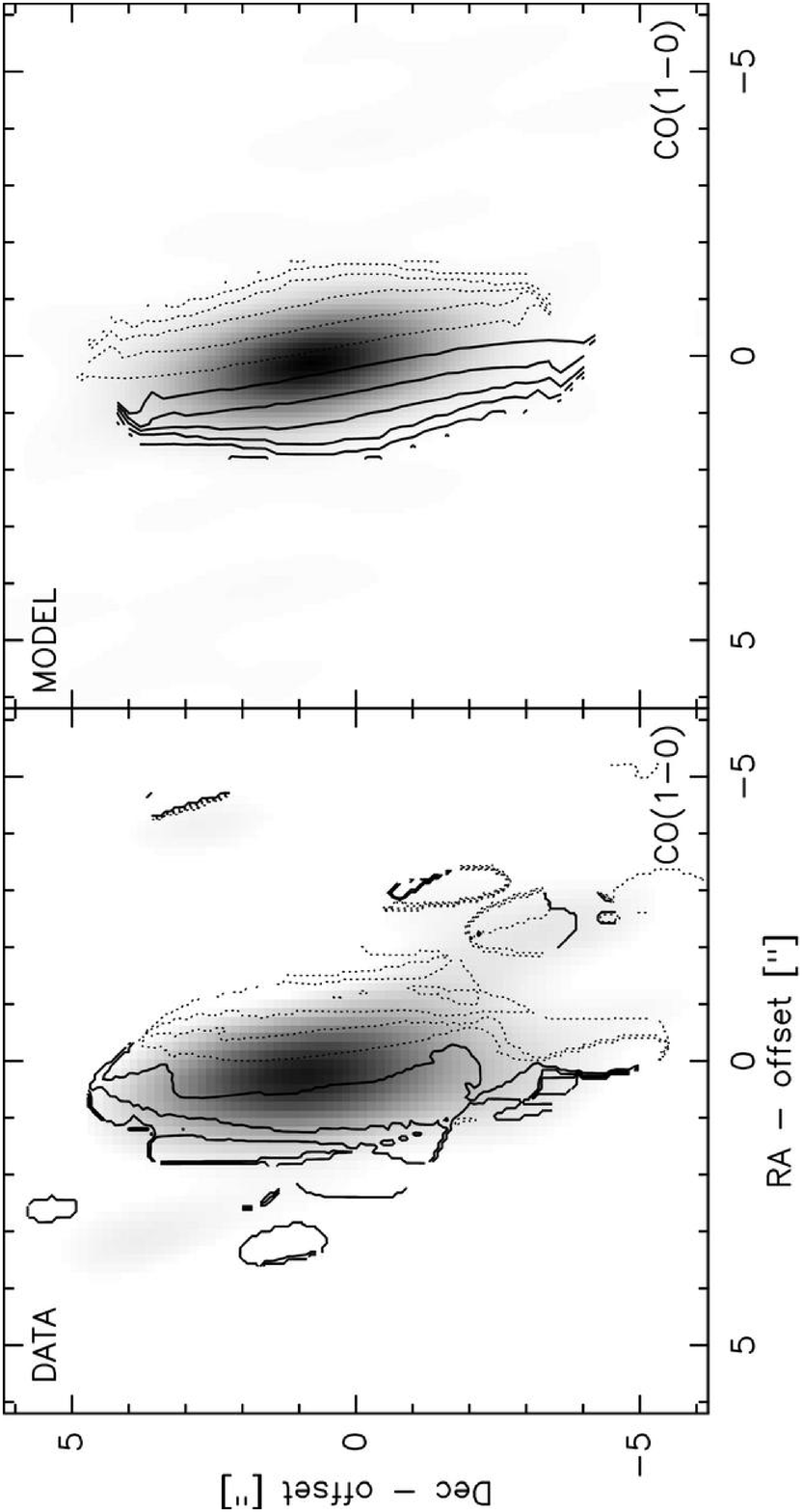}}}
\resizebox{\hsize}{!}{\rotatebox{-90}{\includegraphics{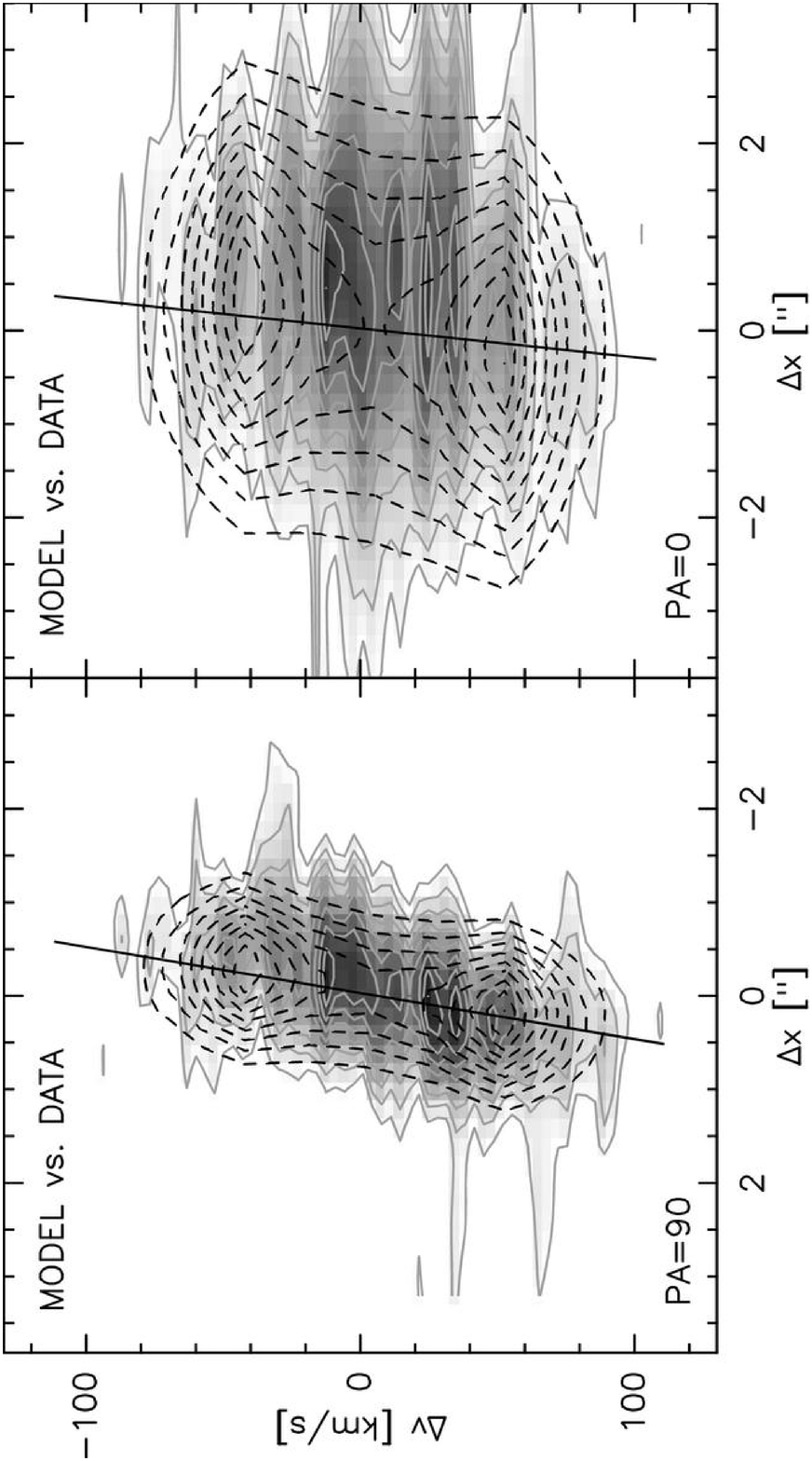}}}
\caption{{\it Upper panel:} Iso-velocity diagrams of the CO(1--0)
emission in HE~1029-1831 ({\it left}) and of a first-order bar model
({\it right}) overlaid to the respective integrated line emission
maps. Increments are again in steps of 10~\kms.{\it Lower panel:}
Position-velocity diagram simulated with a bar model for the emission
in HE~1029-1831 taken along a cut with a PA of 90\degree ({\it left})
and of 0\degree ({\it right}); compare Fig.~\ref{he1029-int-co} \&
\ref{he1029-sli-co}. The dashed contours are superimposed on the
observed data presented in Fig.~\ref{he1029-sli-co}. Increments are in
steps of 10\% from 20\% of the maximum. }
\label{he1029-co-model}
\end{figure}

\section{Gas dynamics}
The kinematics and distribution of the CO emission in HE~1029-1831
suggest a barred potential in this galaxy (compare Roberts et al.\
1979) as visible in the optical (Fischer et al.\ 2006). We have
simulated the gas emission in HE~1029-1831 by adopting the bar
approach from Schinnerer, Eckart \& Tacconi (2000) and Telesco \&
Decher (1988). Instead of computing the gas motions in a given
gravitational potential, the gas is simply split up into several
closed orbits with continous, non-overlapping curves of ellipticities
$\epsilon(r)$ and position angles (pa$(r)$) which are both smooth
functions of the radius. Each ring, i.e., each radius has also an
assigned velocity\footnote{The used code 3DRings is described in more
detail by Schinnerer, Eckart \& Tacconi (2000).}.  For simplification,
we adopt a constant rotation curve and a constant intensity over all
rings. As fixed input parameters, we used the estimated PA of the
emission of $\sim$0-20\degree\, and the inclination of
$\sim$10-30\degree. To find the best-fit model, we compared the
simulated distribution and kinematics for different sets of
$\epsilon(r)$ and pa$(r)$ with those from the observed gas emission.
Fig.~\ref{he1029-co-model} shows the result of these simulations. To
facilitate a comparison to the observed CO(1--0) data, the simulated
CO distribution has been transformed to a uv-table based on the
uv-coverage from the CO(1--0) PdBI observations; this is a more
accurate method than just convolving the model with the beamsize of
the CO(1--0) observations as it considers also spatial filtering
effects due to the discrete uv-coverage. This simplified model already
reproduces the observed emission. The simulated distribution and the
kinematics of the gas agree very well with each other indicating that
the gas properties can be almost completely explained by assuming a
barred potential. However, the asymmetry between the CO(1--0) and
CO(2--1) line cannot be explained by the symmetric bar model.

\section{Summary and Conclusions}
Strong and extended CO(1--0) and CO(2--1) emission is detected in the
barred Seyfert galaxy HE~1029-1831 with the IRAM PdBI and the BIMA
array while no continuum emission was found. Both lines agree well
with each other in distribution, position and velocity width and imply
a total gas mass of $\sim1.2\times10^{10}$\msun. The difference
between the dynamical mass ($\sim$2$\times10^9$\msun, not corrected
for inclination) indicates a low inclination
of\, $\sim$10-30\degree. The CO(1--0) and CO(2--1) emission are
extended along the optical bar. The FWHM of the CO emission is
calculated to be $\sim$(6$\pm$2)kpc. Also, a striking velocity
gradient is found at a position angle of 90\degree, i.e., along the
minor axis of the bar, extending over $\sim1''$ approximately. In the
direction of the bar (PA=0\degree), almost no velocity gradient is
seen.  Adopting the approach from Schinnerer, Eckart \& Tacconi
(2000), the line emission can be simulated with a very simple bar
model explaining the detected distribution and kinematics. Following
Roberts et al.\ (1979), the strong velocity gradient across the bar
(PA=90\degree) can be interpreted as bar-driven inflow of the gas
while the motions along the bar (PA=0\degree) indicate a velocity
dispersion of $\Delta$v=$\pm$80\kms. The line ratio between CO(2--1)
and CO(1--0) is estimated to be $\sim$0.5 over most areas of the bar
although probably still biased by resolution effects. This indicates
subthermally excited and cold gas which is typical for gas in a
bar/disk. The southern increase of the CO(2--1)/CO(1--0) line ratio to
over 1 indicates different excitation conditions which might be caused
by gas compression at the crossing point of the bar and the southern
spiral arm. However, this southern CO component still requires further
support.

\begin{acknowledgements}
      Part of this work was supported by the German
      \emph{Son\-der\-for\-schungs\-be\-reich, SFB\/,} project number
      494. MK was partly funded by a pre-doctoral fellowship of the
      german academic exchange service (DAAD) and the french
      government. This research has made use of the NASA/IPAC
      Extragalactic Database (NED) which is operated by the Jet
      Propulsion Laboratory, California Institute of Technology, under
      contract with the National Aeronautics and Space Administration.
      We thank the anonymous referee for useful and careful comments.
\end{acknowledgements}

\appendix


\begin{thebibliography}{}

\bibitem[2006]{bert06} Bertram, T., Eckart, A., Krips, M.,
Straubmeier, C., Fischer, S., Staguhn, J.~G., New Astronomy Reviews,
Volume 50, Issue 9-10, p.\ 712
\bibitem[2000]{comb00} Combes, F., 2000, ``Dynamics of Galaxies: from
the Early Universe to the Present'', 15th IAP meeting held in Paris,
France, July 9-13, 1999, Eds.: F.\ Combes, G.A.\ Mamon, and V.\
Charmandaris, ASP Conference Series, Vol. 197, ASPC 197, 15
\bibitem[2003]{comb03} Combes, F., Garc\'{i}a-Burillo, S., Boone, F. et
  al.\ 2004, A\&A 414, 857
\bibitem[2004]{comb04} Combes, F., ``The Interplay among Black Holes,
Stars and ISM in Galactic Nuclei'', Proceedings of IAU Symposium,
No. 222. Edited by T. Storchi-Bergmann, L.C. Ho, and Henrique
R. Schmitt., Cambridge, UK: Cambridge University Press, 2004.,
p.383-388
\bibitem[2003]{cont03}	Contini, M., \& Contini, T., 2003, MNRAS, 342, 299
\bibitem[1998]{down98} Downes, D., \& Solomon, P. M. 1998, ApJ, 507, 615
\bibitem[2006]{fisc06} Fischer, S., Iserlohe, C., Zuther, J., Bertram, T.,
Straubmeier, C., Schoedel, R., Eckart A., A\&A, 2006, 452, 827
\bibitem[2003]{garca} Garc\'{\i}a-Burillo, S., Combes, F., Eckart, A.,
et al.\ 2003a, in Active Galactic Nuclei: From Central Engine to Host
Galaxy, ed. S. Collin, F.Combes \& I.Shlohsman, ASP Conf.Ser.\ 290,
423
\bibitem[2003]{garcb} Garc\'{\i}a-Burillo, S., Combes, F., Hunt, L.K.,
Boone, F. et al.\ 2003b, A\&A 407, 485
\bibitem[2005]{garc05} Garc\'{\i}a-Burillo, S.; Combes, F.;
Schinnerer, E.; Boone, F.; Hunt, L. K., 2005, A\&A, 441, 1011
\bibitem[2003]{kal} Kaldare, R., Colless, M., Raychaudhury, S., 
Peterson, B.A., 2003, MNRAS, 339, 652
\bibitem[2001]{kew} Kewley, L. J., Heisler, C. A., Dopita, M. A. \and
Lumsden, S.\ 2001, ApJS, 132, 37
\bibitem[2005]{kri} Krips,M., Eckart, A., Neri, R., Pott, J.-U., Leon,
S., Combes, F., Garc\'{\i}a-Burillo, S., Hunt, L.K., Baker, A.J.,
Tacconi, L.J., Englmaier, P., Schinnerer, E. \and Boone, F.\ 2005,
A\&A, 442, 479
\bibitem[1979]{robe79} Roberts, W.W., Huntley, J.M., and van Albada, G.D.,
1979, ApJ, 233, 67
\bibitem[2000]{schi00} Schinnerer, E., Eckart, A., Tacconi, L. J., 2000,
ApJ, 533, 826
\bibitem[1991]{solo91} Solomon, P. M., \& Barrett, J. W., 1991,
``Dynamics of Galaxies and Their Molecular Cloud Distributions'',
Proceedings of the 146th Symposium of the International Astronomical
Union, held in Paris, France, June 4-9, 1990, edited by F.\ Combes and
F.\ Casoli, International Astronomical Union, Symposium no.146, Kluwer
Academic Publishers, Dordrecht, 1991., p.235
\bibitem[1988]{tele88} Telesco, C. M., \& Decher, R., 1988, ApJ, 334,
573

\end{thebibliography}
\end{document}